\documentclass[prd,floats,floatfix,showkeys,nofootinbib]{revtex4}

\pdfoutput=1
\usepackage{graphicx}
\usepackage{dcolumn}
\usepackage{bm}
\usepackage{graphics}
\usepackage{slashed}
\usepackage{amssymb}
\usepackage{natbib}
\usepackage{amsmath}
\usepackage{amsthm}
\usepackage{url}
\usepackage{color}
\usepackage{xcolor}
\newcommand{\fracc}[2]{\frac{\textstyle{#1}}{\textstyle{#2}}}

\begin{document}

\title{Scalar perturbations in a Friedmann-like metric with non-null Weyl tensor}

\author{G. B. Santos}\email{grasiele.dossantos@icranet.org}
\author{E. Bittencourt}\email{eduardo.bittencourt@icranet.org}
\affiliation{Dipartimento di Fisica, Universit\`a ``La Sapienza'', P.le A. Moro 2, 00185 Roma, Italia and CAPES Foundation, Ministry of Education of Brazil, Brasilia, Brazil}
\author{J. M. Salim}\email{jsalim@cbpf.br}\affiliation{Instituto de Cosmologia Relatividade e Astrofisica ICRA -
CBPF\\ Rua Doutor Xavier Sigaud, 150, CEP 22290-180, Rio de Janeiro,
Brazil}

\keywords{cosmological perturbation theory,primordial magnetic fields,physics of the early universe}
\date{\today}

\begin{abstract}
In a previous work the authors have solved the Einstein equations of General Relativity for a class of metrics with constant spatial curvature, where it was found a non vanishing Weyl tensor in the presence of a primordial magnetic field with an anisotropic pressure component. Here, we perform the perturbative analysis of this model in order to study the gravitational stability under linear scalar perturbations. For this purpose, we take the Quasi-Maxwellian formalism of General Relativity as our framework, which offers a naturally covariant and gauge-invariant approach to deal with perturbations that are directly linked to observational quantities. We then compare this scenario with the perturbed dust-dominated Friedmann model emphasizing how the growth of density perturbations are enhanced in our case.
\end{abstract}

\maketitle

\section{Introduction}
There are basically two approaches to perturbation theory in general relativity (GR): the Lifshitz's method \cite{lifshitz,lifs_khal} which was rigorously improved by the work of Bardeen \cite{bardeen} and others \cite{kodama, mukh92} and the covariant method given by the Quasi-Maxwellian (QM) equations \cite{jek}, which was first proposed by Hawking \cite{hawking_66} and improved by Olson \cite{olson}, Ellis {\it et al} \cite{ellis_bruni,ellis} and Novello {\it et al} \cite{novello,nbs}. The former includes perturbations of non-observable quantities such as $\delta g_{\mu\nu}$ and it mixes true perturbations with coordinate transformations coming from the gauge freedom due to the arbitrariness of the correspondence between a fictitious background space-time and the physical inhomogeneous one.
In Bardeen's approach gauge-invariant quantities are constructed as combinations of the perturbations of the metric and of the matter content, but no clear physical interpretation can be given to those variables.

On the other hand, the QM equations offer a natural covariant approach to perturbation theory, in which it is possible to find suitable gauge-invariant variables that are directly linked to observational quantities, thus making the physical interpretation clear and providing an elegant description. In this approach, it is possible to find a set of perturbed quantities which are considered as ``good'' ones as their unperturbed counterparts are null in the background and, therefore, Stewart \& Walker's lemma \cite{stewart} ensures that the associated perturbed quantities are gauge-invariant. In \cite{novello} for instance this method was applied to the Friedmann-Lema\^itre-Robertson-Walker (FLRW) models in order to find the minimal closed set of variables necessary to the perturbative analysis, in terms of which all other quantities could be written. Remarkably, this minimal set contained only two of the full set of perturbation variables providing a planar dynamical system, from which all the other quantities could be expressed. The relation between the covariant formalism and Bardeen's approach have been clarified in Refs. \cite{hwang, bruni92}.

In our case, the presence of an anisotropic pressure due to a primordial magnetic field (PMF) removes some symmetries of the FLRW models and makes the analysis a bit more complex. We show that it is possible to construct a minimal set consisting of three gauge-invariant variables contained in the full set we have at our disposal. The latter includes the fractional gradient of the energy density and the gradient of the expansion coefficient, both orthogonal to the fluid flow, which were introduced by Ellis and Bruni \cite{ellis_bruni}. In the special case of large wavelengths, when the equations for the perturbations reduces to a planar system, it is possible to construct a second order equation from which we can analyze the growth of inhomogeneities.

This work is organized as follows. In section \ref{notation} we give the main definitions needed for the description of perturbations in the QM approach. In section \ref{background} we briefly introduce the background model and its important features. In section \ref{basis} we construct the complete basis for the perturbations and derive from them the necessary quantities for the subsequent analysis. Afterwards, in section \ref{set}, we present the set of variables needed to describe the perturbations and their corresponding evolution equations. In section \ref{pert} we perform the perturbative analysis focusing on the large wavelength behavior. Then, we compare our results with the standard FLRW model in order to verify how the growth of perturbations are modified in this new scenario and if the anisotropic pressure could play a role similar to a dark matter component in the standard cosmological model. The full set of the Quasi-Maxwellian equations, the conservation laws for the matter content and the evolution equations for the kinematical quantities are presented in the Appendix.

\section{Definitions and Notation}
\label{notation}

We consider the standard definitions given in \cite{nbs} to introduce the QM approach to General Relativity, except for some conventions (for details see the Appendix). The Weyl conformal tensor is defined as\footnote{Greek indices run from 0 to 3 and latin indices run from 1 to 3.}
\begin{equation}
\nonumber
W_{\alpha\beta\mu\nu}\doteq R_{\alpha\beta\mu\nu} - M_{\alpha\beta\mu\nu} + \fracc{1}{6} R g_{\alpha\beta\mu\nu},
\end{equation}
where the auxiliary tensors are
\begin{equation}
\nonumber
2M_{\alpha\beta\mu\nu} \doteq R_{\alpha\mu}g_{\beta\nu} + R_{\beta\nu}g_{\alpha\mu} - R_{\alpha\nu}g_{\beta\mu} - R_{\beta\mu}g_{\alpha\nu}
\end{equation}
and
\begin{equation}
\nonumber
g_{\alpha\beta\mu\nu}\doteq g_{\alpha\mu}g_{\beta\nu}-g_{\alpha\nu}g_{\beta\mu}.
\end{equation}

The Weyl tensor can be separated into its electric and magnetic parts, defined as
\begin{equation}
\nonumber
\begin{array}{lcl}
E_{\alpha\beta}&\doteq&-W_{\alpha\mu\beta\nu}V^{\mu}V^{\nu},\\[2ex]
H_{\alpha\beta}&\doteq&-^{*}W_{\alpha\mu\beta\nu}V^{\mu}V^{\nu},
\end{array}
\end{equation}
where $^{*}W_{\alpha\mu\beta\nu}$ is the dual Weyl tensor constructed with the skew-symmetric Levi-Civita tensor. The tensors $E_{\alpha\beta}$ and $H_{\alpha\beta}$ are symmetric, traceless and belong to the 3-dimensional space, orthogonal to a class of observers with four-velocity $V^\mu$.

The metric $g_{\mu\nu}$ and the normalized vector field $V^\mu$ (tangent to a timelike congruence of curves $\Gamma$) induce a projector $h_{\mu\nu}$ which splits tensors in terms of quantities defined along $\Gamma$ plus quantities defined on the three-dimensional space orthogonal to $V^\mu$. It is expressed as
\begin{equation}
\label{proj}
h_\mu{}^\nu\doteq\delta_\mu{}^\nu-V_\mu V^\nu.
\end{equation}

We deal here with a Friedmann type of metric which, as such, is written as
\begin{equation}
\label{gen_ds2}
ds^2=dt^2+h_{ij}dx^idx^j,
\end{equation}
where $h_{ij}=-a^2(t)\gamma_{ij}(x^k)$ and the modification with respect to the Friedmann models is encoded in the $\gamma_{ij}$ part, as we shall see.

We introduce the so-called kinematical quantities through the decomposition of the covariant derivative of $V^{\mu}$ into its irreducible parts:
\begin{equation}
V_{\mu;\nu} \doteq \sigma_{\mu\nu} + \omega_{\mu\nu} + \fracc{1}{3}\theta h_{\mu\nu} + a_{\mu}V_{\nu},
\nonumber
\end{equation}
where the expansion coefficient is
$$\theta\doteq V^{\mu}{}_{;\mu},$$
the acceleration is
$$a^{\mu}\doteq V^{\mu}{}_{;\nu}V^{\nu},$$
the traceless and symmetric shear tensor is
$$\sigma_{\mu\nu} \doteq \frac{1}{2} h_{\mu}{}^{\alpha} h_{\nu}{}^{\beta} V_{(\alpha;\beta)} - \frac{\theta}{3} h_{\mu\nu},$$
and the skew-symmetric vorticity tensor is
$$\omega_{\mu\nu}\doteq\frac{1}{2} h_{\mu}{}^{\alpha} h_{\nu}{}^{\beta} V_{[\alpha;\beta]},$$
where $()$ means symmetrization and $[\,]$ means anti-symmetrization.



\section{Friedmann equations in the presence of an anisotropic pressure}
\label{background}

In this section, we briefly review the recent proposal \cite{bitt_salim} in which it is assumed a Friedmann-like geometry
\begin{equation}
\label{fried}
ds^2=dt^2-a^2(t)[d\chi^2+\sigma^2(\chi)d\Omega^2],
\end{equation}
where $t$ represents the cosmic time, $a(t)$ is the scale factor and $\sigma(\chi)$ is an arbitrary function.

We consider as source the linear Lagrangian of Maxwell's theory of electromagnetism
$$L=-\fracc{1}{4}\,F,$$
where $F\equiv\,F^{\mu\nu} F_{\mu\nu} = 2 (B^2-E^2)$.
The energy-momentum tensor corresponding to this Lagrangian is
$$T_{\mu\nu}=F_{\mu}{}^{\alpha}F_{\alpha\nu}-Lg_{\mu\nu},$$
whose decomposition into irreducible parts with respect to a normalized time-like vector field $V^{\mu}$
yields
\begin{equation}
\label{rho_p}
\rho = \frac{1}{2}(E^2+B^2), \hspace{.5cm} p =\frac{\rho}{3},
\end{equation}
\begin{equation}
\label{q}
q^{\alpha} = \eta^{\alpha}{}_{\beta\mu\nu} V^{\beta} E^{\mu} B^{\nu}
\end{equation}
and
\begin{equation}
\label{pi_em}
\pi_{\mu\nu} = -E_{\mu} E_{\nu} -B_{\mu} B_{\nu} - \frac{1}{3} (E^2+B^2) h_{\mu\nu},
\end{equation}
where $\rho$ is the energy density, $p$ is the isotropic pressure, $q^{\alpha}$ is the heat flux and $\pi_{\mu\nu}$ is the anisotropic pressure. It is convenient to assume the cosmic observer $V^{\mu}=\delta^{\mu}_0$. Due to the special symmetries of the metric\ (\ref{fried}), the electromagnetic field can be considered as source of the gravitational field only if an averaging process is performed (cf.\ Ref.\ \cite{tolman2,hind}). The standard way to compute the volumetric spatial average of an arbitrary quantity X at the instant $t=t_0$ is defined by

\begin{equation}
\label{av_def}
\overline{X}\,\equiv \lim_{V \rightarrow V_0} \fracc{1}{V} \int X\sqrt{-g}d^3x^i,
\end{equation}
where we denote $V\equiv\int \sqrt{-g}d^3x^i$ and $V_0$ is a sufficiently large time dependent spatial volume. Therefore, the first and second moments of the electric $E_i$ and magnetic $H_i$ fields are usually given by the so-called Tolman relations:

\begin{equation}
\label{rel_tol_1}
\overline{E_i}=0, \hspace{.3cm} \overline{H_i}=0, \hspace{.3cm} \overline{E_iH_j}=0,
\end{equation}

\begin{equation}
\label{rel_tol_2}
\overline{E^iE_j}=-\frac{1}{3}E^2h^i{}_j,
\end{equation}

\begin{equation}
\label{rel_tol_3}
\overline{B^iB_j}= -\frac{1}{3}B^2h^i{}_j.
\end{equation}
We keep only the magnetic component of this mean field due to the high conductivity of the primordial plasma. Moreover, we do not impose homogeneity of the space-time but keep only the isotropy. Therefore, part of the Tolman relations are preserved [Eq.\ (\ref{rel_tol_1})]. However, our choice leads us to slightly modify the Tolman relation concerning the second moment of the magnetic field (\ref{rel_tol_3}), as follows

\begin{equation}
\label{rel_tol_3_mod}
\overline{B^iB_j}=-\frac{1}{3}B^2h^i_j-\pi^{i}{}_{j},
\end{equation}
where we introduce an arbitrary traceless matrix $\pi^{i}{}_{j}$ that will be identified to an anisotropic pressure term, through Eq.\ (\ref{pi_em}).

In the case of constant spatial curvature $^{(3)}R$, the time evolution of this cosmological model is driven by the usual Friedmann equations and the anisotropic pressure produces a non-vanishing Weyl tensor. The anisotropic pressure components are found to be

\begin{equation}
\label{pi}
\pi^{2}{}_{2}=\pi^{3}{}_{3},\hspace{1cm}\pi^{1}{}_{1}=-2\pi^{2}{}_{2}=\fracc{2k}{a^2\sigma^3},
\end{equation}
where $k$ is a constant\footnote{It is actually a constant coming from the integration of Einstein equations, see Ref. \cite{bitt_salim} for details.}. The corresponding Riemannian tensor $^{(3)}R_{\alpha\beta\mu\nu}$ of the spatial hypersurface can be written as
\begin{displaymath}
^{(3)}R_{\alpha\beta\mu\nu} = \fracc{\epsilon}{a^2} h_{\alpha\beta\mu\nu} - \pi_{\alpha[\mu}h_{\nu]\beta} + \pi_{\beta[\mu}h_{\nu]\alpha},
\end{displaymath}
where $h_{\alpha\beta\mu\nu} \equiv h_{\alpha\mu}h_{\beta\nu} - h_{\alpha\nu}h_{\beta\mu}$. This geometry is no longer maximally symmetric due to the presence of the last two terms.



Finally, making the coordinate transformation given by $r=\sigma(\chi)$, the line element\ (\ref{fried}) becomes

\begin{equation}
\label{fried_r_pi}
ds^2=dt^2-a^2(t)\left(\fracc{dr^2}{1-\epsilon r^2-\frac{2k}{r}}+r^2d\Omega^2\right).
\end{equation}

The corresponding QM equations of this solution are (see details in Appendix):

\begin{subequations}
\label{qm_gfried}
\begin{eqnarray}
&&\dot\theta+\fracc{\theta^2}{3}=-\fracc{1}{2}(\rho+3p),\label{qm_gfried1}\\[2ex]
&&\dot\rho+(\rho+p)\,\theta=0,\label{qm_gfried2}\\[2ex]
&&E_{\mu\nu}=-\fracc{1}{2}\pi_{\mu\nu},\label{qm_gfried3}\\[2ex]
&&E^{\alpha}{}_{\mu;\alpha}=0,\label{qm_gfried4}\\[2ex]
&&h^{\epsilon}{}_{\mu}h^{\nu}{}_{\lambda}\dot E^{\mu}{}_{\nu} + \frac{2}{3}\,\theta\, E^{\epsilon}{}_{\lambda} =0.\label{qm_gfried5}
\end{eqnarray}
\end{subequations}
Eqs.\ (\ref{qm_gfried1}) and (\ref{qm_gfried2}) correspond to the usual Friedmann equations

\begin{equation}
\label{gr_fried_solved}
H^2+\fracc{\epsilon}{a^2}=\fracc{\rho}{3},\hspace{1cm}\fracc{\ddot a}{a}=-\fracc{1}{6}(3\gamma-2)\rho,
\end{equation}
where $H\equiv\dot a/a$ is the Hubble parameter. Eq.\ (\ref{qm_gfried3}) yields immediately the electric part of the Weyl tensor in terms of the anisotropic pressure. Eq.\ (\ref{qm_gfried4}) represents the compatibility condition of the Einstein equations with the (constant) spatial curvature and Eq.\ (\ref{qm_gfried5}) provides the time evolution of $E_{\mu\nu}$. Note that these equations do not imply that the Weyl tensor is identically zero. Once Eqs.\ (\ref{qm_gfried1}-\ref{qm_gfried2}) are decoupled from Eqs.\ (\ref{qm_gfried3}-\ref{qm_gfried4}) we see that this model can be extended to any equation of state (EOS) of the form $p=(\gamma-1)\rho$, which is also valid for a mixture of non-interacting fluids.

From a straightforward calculation, the electric part of Weyl tensor for the cosmic observer reads

\begin{equation}
\label{el_weyl_r}
[E^i{}_j]=\fracc{k}{a^2r^3}
\left(
\begin{array}{ccc}
-1&0&0\\[.5ex]
0&\frac{1}{2}&0\\[.5ex]
0&0&\frac{1}{2}
\end{array}
\right).
\end{equation}
This equation is very similar to the Newtonian tidal forces multiplied by a time dependent function. We conclude that the Friedmann equations modify the Weyl tensor through the solution for $a(t)$, but the Weyl tensor does not change the Friedmann equations.

The remarkable exact relation\ (\ref{qm_gfried3}), which was not highlighted in the literature before\footnote{In fact, this solution was briefly presented in Lema\^itre's work \cite{lem33} and its mathematical features were more recently displayed by McManus and Coley \cite{coley}, but its full analysis and the issues raised in \cite{bitt_salim} were not envisioned in those works.}, informs us that the dissipative process to which the fluid is subjected is compensated in such a way that the tidal forces of the constituting microscopic parts of the fluid produce a reversal effect. This fact is confirmed in the geodesic deviation of the fluid flow which remains undistorted by these quantities. We now proceed with the set up of the framework for treating the perturbed version of this model.


\section{Construction of the basis}
\label{basis}
Since the seminal Lifshitz paper \cite{lifshitz} and the improvements displayed by Lifshitz and Khalatinikov \cite{lifs_khal}, it has been useful to expand all perturbed quantities in the spherical harmonics basis. Once we are interested only in the linear regime of the perturbations, the Laplace-Beltrami operator defined at the 3-hypersurface orthogonal to $V^{\mu}$ provides a complete set of functions which expand all perturbed quantities lying on this surface. This is also achieved due to the isotropy of the metric\ (\ref{fried_r_pi}), which allows the separation of variables transforming the eigenvalue problem in a set of ordinary differential equations to be solved.

It suffices for our purposes to take into account only the spatial scalar harmonic functions $Q(x^k)$ and its derived vector and tensor quantities
\begin{eqnarray}
&&Q_{i}\doteq Q_{,i},\nonumber\\
&&Q_{ij}\doteq Q_{,i||j}=Q_{,i;j}.\nonumber
\end{eqnarray}
For a while, it is necessary to distinguish covariant derivative in the 4-dimensional space-time by the symbol (;) and the derivative on the spatial hypersurface by (${}_{\|}$).

The functions $Q(x^k)$ obey the following eigenvalue equation in the 3-dimensional background space for each mode
\begin{equation}
 \nabla^2Q_{(m)}=m^2Q_{(m)},
\end{equation}
where $m$ is a constant (the wave number) and
\begin{equation}
\label{lap_bel}
\nabla^2Q\doteq \gamma^{ij}Q_{,i||j}=\gamma^{ij}Q_{,i;j},
\end{equation}
defines the 3-dimensional Laplace-Beltrami operator. The general solution of this equation is given by
$$Q(r,\theta,\phi)=\sum_{p,n}R(r)Y^n_p(\theta,\phi),$$
where $Y^n_p(\theta,\phi)$, with $p=0,1,2,...$ and $n=-p,...,p$, are the spherical harmonics and $R(r)$ satisfies

\begin{equation}
\label{r_eq}
\fracc{d}{dr}\left[r(r-2k-\epsilon r^3)\fracc{dR}{dr}\right] - [m^2r^2+p(p+1)]R=0.
\end{equation}
Note that only the complete integration of the above differential equation can inform the range of values allowed for $m^2$. However, it doesn't have a known analytic solution for arbitrary values of the parameters, particularly, $\epsilon=0$ and $r>2k$ which we are interested in.

We define the traceless operator
\begin{equation}
\hat Q_{ij}=\frac{1}{m^2}Q_{ij}-\frac{1}{3}Q\gamma_{ij},
\end{equation}
and its divergence can be computed yielding
\begin{equation}
\hat Q^{j}{}_{i||j}= 2\left(\frac{1}{3}-\frac{\epsilon}{m^2}\right) Q_i-\frac{\pi_{ij}}{m^2} Q^j.
\end{equation}
In the expression above the Einstein equations were used and $\pi_{ij}$ is a tensor involving only the spatial coordinates.

In particular, all possible projections of the anisotropic pressure\footnote{In the metric under consideration, the anisotropic pressure is the only nonzero tensorial quantity on the background. The electric part of the Weyl tensor, which is also nonzero, is determined by $\pi_{\mu\nu}$.} on the basis components can be rewritten in terms of this basis, if we assume that it is a complete one. Therefore, we write

\begin{equation}
\label{scalar_exp}
\pi_{ij} \hat Q^{ij}_{(m)}=\sum_la_{l(m)}Q_{(l)},
\end{equation}
\begin{equation}
\label{vector_exp}
\pi_{ij} Q^j_{(m)}=\sum_lb_{l(m)}Q_{i(l)},
\end{equation}
and
\begin{equation}
\label{tensor_exp}
\frac{1}{2}\pi_{k(i} \hat Q_{j)}{}^{k}{}_{(m)}=\sum_lc_{l(m)}\hat Q_{ij(l)}+\fracc{\gamma_{ij}}{3}\sum_la_{l(m)}Q_{(l)},
\end{equation}
where the coefficients $a_{l(m)}$, $b_{l(m)}$ and $c_{l(m)}$ are constants for each of the modes $m$ and $l$. Note that the summation symbol should be replaced by an integration symbol depending on the spectrum of the Laplace operator.

For our purposes, it is not important the explicit form of the scalar functions, but instead only the fact that such functions form a complete set with which it is possible to expand all perturbed quantities. Therefore, assuming a small deviation of the metric given in (\ref{fried}) when compared to FLRW, the quantities
$$A_{(m)}\doteq\sum_la_{l(m)}, \quad B_{(m)}\doteq\sum_lb_{l(m)}, \quad C_{(m)}\doteq\sum_lc_{l(m)},$$
should be necessarily bounded, but they are not arbitrarily small. Note that $A$ and $B$ are not independent quantities: from Eqs. (\ref{scalar_exp}) and (\ref{vector_exp}) we have $$A_{(m)}=\frac{1}{m^2}\sum_l b_{l(m)}l^2.$$
Note also that once the free parameter $k$ of the metric\ (\ref{fried_r_pi}) is fixed, and this should be achieved through observations, all these quantities should be completely determined.\\


\section{Gauge-invariant variables and unperturbed equations}
\label{set}
In this section we search for a set of gauge-invariant variables needed to perform the perturbative analysis and their correspondent evolution equations.
According to the evolution equation for the shear tensor (\ref{evol_quant_cine2}), we have
\begin{equation}
X_{\mu\nu}\doteq E_{\mu\nu}+\frac{1}{2}\pi_{\mu\nu},
\end{equation}
which is a good variable as it is null in the background, hence a perturbation on it yields a true physical perturbation. To this variable we add the shear $\sigma_{\mu\nu}$ itself, which is also null in the background. Following \cite{ellis_bruni}, we also consider as good variables the comoving fractional energy density gradient

\begin{equation}
\chi_\alpha\doteq a(t)\frac{h_\alpha{}^\nu\rho_{,\nu}}{\rho},
\end{equation}
and the comoving gradient of the expansion coefficient
\begin{equation}
 Z_\alpha\doteq a(t)h_\alpha{}^\nu\theta_{,\nu}.
\end{equation}

To this set of variables, we can still add other quantities that are null on the background: the acceleration $a_\mu$ and the divergence of the anisotropic pressure $I_\mu\equiv h_\mu{}^\epsilon \pi_\epsilon{}^\nu{}_{;\nu}$.

The unperturbed equations for the above variables follow from the evolution equations (\ref{quase_max4}), (\ref{expl_proj_conserv_mom_eneg1}), (\ref{evol_quant_cine1}) and (\ref{evol_quant_cine2}) and are written as

\begin{equation}
\label{X}
h_\mu{}^\epsilon h_\nu{}^\lambda \dot X_{\epsilon\lambda} =- \theta X_{\mu\nu} - \frac{1}{2}\pi_{\alpha(\mu}\sigma_{\nu)}{}^\alpha + \frac{1}{3}\pi_{\alpha\beta}\sigma^{\alpha\beta}h_{\mu\nu}  -\frac{1}{2}(\rho_t+p_t)\sigma_{\mu\nu}+D_{\mu\nu},
\end{equation}
\begin{equation}
\label{chi}
h_\mu{}^\lambda \dot \chi_\lambda=-\gamma_{ef}Z_\mu + \frac{a}{\rho}h_\mu{}^\lambda(\pi^{\alpha\beta}\sigma_{\alpha\beta})_{,\lambda} + a\,\gamma_{ef}\,\theta\,a_\mu,
\end{equation}
\begin{equation}
\label{theta}
h_\mu{}^\alpha \dot Z_\alpha=-a\,\dot\theta\, a_\mu-\frac{2\theta}{3a} Z_\mu+h_\mu{}^\alpha a^\nu{}_{;\nu;\alpha}-\frac{1}{2}(3\gamma_{ef}-2)\frac{\rho}{a}\chi_\mu,
\end{equation}
\begin{equation}
\label{sigma}
h_{\mu}{}^{\epsilon}h_{\nu}{}^{\lambda}\dot\sigma_{\epsilon\lambda} =- \fracc{1}{3}h_{\mu\nu}a^{\lambda}{}_{;\lambda} - a_{\mu}a_{\nu}+\fracc{1}{2}h_{\mu}{}^{\epsilon}h_{\nu}{}^{\lambda}a_{(\epsilon;\lambda)} - \fracc{2}{3}\theta\sigma_{\mu\nu}-X_{\mu\nu},
\end{equation}
where $D_{\mu\nu}\doteq\frac{2}{3}\theta\pi_{\mu\nu}+h_\mu{}^\epsilon h_\nu{}^\lambda \dot \pi_{\epsilon\lambda}$ and terms that would generate second order perturbations have already been discarded. The quantities $\rho_t$ and $p_t$ represent the total energy density and pressure, respectively, of a non-interacting mixture of simple fluids. They are related through the equation of state $p_t=(\gamma_{ef}-1)\rho_t$ where we have introduced an effective EOS parameter \cite{ellis_bruni} which, in the case of a universe filled with dust and radiation, is given by\footnote{The EOS parameter in this case depend on time, but $\dot\gamma_{ef}\approx 0$ and we can always assume that it has the value of the dominating fluid, cf.\ \cite{ellis_bruni}.}
$$\gamma_{ef}=\frac{\rho_d+4/3\rho_r}{\rho_d+\rho_r}.$$
The frame chosen is such that there's no energy flux, that is, $q^\alpha=0$. To these equations we must add the non-trivial constraint equations (\ref{quase_max1}), (\ref{expl_proj_conserv_mom_eneg2}) and (\ref{eq_vinc_quant_cine1}).

The term $D_{\mu\nu}$ in Eq. (\ref{X}) is determined by the causal thermodynamical relation \cite{israel} restricted by the symmetries of (\ref{fried}). It reads
\begin{equation}
\label{gen_israel_eq}
\tau\dot\pi_{\mu\nu}+\pi_{\mu\nu}=\xi\sigma_{\mu\nu},
\end{equation}
where $\tau$ is the relaxation time and $\xi$ is the viscosity parameter. This equation is valid only in the linear perturbation regime and was derived guided by the principle that the entropy flux is to be strictly local, that is, independent of gradients of the energy-momentum tensor and of the particle flux vector. In our case, in the background we have $$\tau\dot\pi_{\mu\nu}+\pi_{\mu\nu}=0$$ such that relations (\ref{pi}) give the relaxation time
$$\tau=\frac{3}{2\theta}.$$
The relation $\tau\propto 1/\theta$ was already considered in previous works in different contexts \cite{longa, novelloduque}.

\section{Perturbation Theory}
\label{pert}

From the original QM equations we have obtained equations for a set of suitable variables as described in the last section and now we aim to construct the perturbed version of them up to first order scalar perturbations. Similar analyses have been performed in Refs \cite{barrow98,tsagas00} where they have considered the contribution of anisotropic stresses on top of the usual perfect fluid content to the evolution of perturbations of a standard FLRW universe using the covariant approach. In particular, in \cite{tsagas00}, this anisotropic stress is due specifically to a vector magnetic field that should be small in order not to break isotropy. In both cases, terms that couple the anisotropic pressure with shear have been considered as second order as they indeed are in a FLRW background. In our case, however, as seen in section \ref{background}, we restrict ourselves to an anisotropic pressure component non-null at the background caused by random PMF--which has no vector character--and whose effect cannot be perceived at the zeroth order due to a compensation by the electric part of the Weyl tensor. In short, $\pi_{\mu\nu}\neq 0$ at the background, which turns out to be a modified FLRW, and that's the reason why we have to consider terms like $\pi_{\alpha\beta}\sigma^{\alpha\beta}$ as first order ones.

In what follows we present the full set of equations for perturbations at linear order. In particular, in the long wavelength regime, we get a system that is closed in only two variables allowing to perform a simpler phenomenological analysis and to obtain a second order equation for the evolution of such perturbations. Then, a comparison with the standard cosmological scenario in this regime is made.

\subsection{General Case}

We now consider the perturbative analysis associated to the variables previously defined. For instance, the perturbed version of Eq.\ (\ref{X}) is given by
\begin{equation}
\label{deltaX}
h_\mu^\epsilon h_\nu^\lambda \dot{\delta X_{\epsilon\lambda}} + \theta\delta X_{\mu\nu} + \fracc{1}{2}\pi_{\alpha(\mu}\delta\sigma_{\nu)}{}^\alpha - \fracc{1}{3}\pi_{\alpha\beta}\delta\sigma^{\alpha\beta}h_{\mu\nu} = -\fracc{1}{2}\gamma_{ef}\rho_t\delta\sigma_{\mu\nu}+\delta D_{\mu\nu},
\end{equation}
where $\delta D_{\mu\nu}=\xi\theta\delta\sigma_{\mu\nu}$ is given through the perturbed thermodynamical relation (\ref{gen_israel_eq}).

We use the complete basis introduced in Sec.\ \ref{basis} in order to expand the perturbations. For the perturbed equation for the quantity $X_{\epsilon\lambda}$ we set $\delta X_{ij}=X(t)\hat Q_{ij}$ and, for the shear, $\delta \sigma_{ij}=\sigma(t)\hat Q_{ij}$. In order to perform the expansion we need to consider terms like
\begin{eqnarray}
h_{\mu\nu}\pi_{\alpha\beta}\delta\sigma^{\alpha\beta}&=&h_{\mu\nu}\pi_{\alpha\beta}h^{\alpha\gamma}h^{\beta\delta}\delta\sigma_{\gamma\delta}\nonumber\\
&=&\frac{\sigma(t)}{a^4}Ah_{\mu\nu}Q,
\end{eqnarray}
where we have used relation (\ref{scalar_exp}). Remember that $A$ is constant in time but depends on the entire spectrum of wave numbers\footnote{Note that every quantity carries an index associated to its wavenumber, as usual when doing this kind of expansion. Thus we have omitted the subscript $(m)$ in $A$ for consistency of notation, as we have omitted it in the other quantities as well.}. We shall also consider relations (\ref{scalar_exp}) and (\ref{tensor_exp}) to write
\begin{equation}
\fracc{1}{2} \pi_{\alpha(\nu}\delta\sigma_{\mu)}{}^{\alpha} - \fracc{1}{3}\pi_{\alpha\beta}\delta\sigma^{\alpha\beta}h_{\mu\nu} = -\fracc{\sigma(t)}{a^2}C\hat Q_{\mu\nu}.
\end{equation}
which will be used in the expansion of Eq. (\ref{deltaX}).

In an analogous way, for the remaining variables we consider the expansions $\delta \chi_{i}=\tilde{\chi}(t) Q_{i}$, $\delta Z_{i}=Z(t) Q_{i}$, $\delta a_{i}=\psi(t) Q_{i}$ and $\delta I_{i}=I(t) Q_{i}$. The perturbed equations then result
\begin{equation}
\label{xt}
\dot X + \theta X + \left(-\fracc{C}{a^2} + \frac{1}{2}\gamma_{ef}\rho - \xi\theta\right)\sigma=0,
\end{equation}
\begin{equation}
\label{sigmat}
\dot \sigma-m^2\psi+X=0,
\end{equation}

\begin{equation}
\label{zt}
\dot Z + \left(a\,\dot\theta-\fracc{m^2}{a^2}\right)\psi+\frac{2\theta}{3a} Z + \frac{1}{2}(3\gamma_{ef}-1)\rho_t\tilde{\chi} = 0,
\end{equation}

\begin{equation}
\label{chit}
\dot{\tilde{\chi}} + \gamma_{ef}Z - \frac{1}{a^3}\frac{A}{\rho_t}\sigma - a\gamma_{ef}\theta\psi=0.
\end{equation}

The constraint equations yield
\begin{eqnarray}
\gamma_{ef}\psi+\frac{I}{\rho}-(\gamma_{ef}-1)\frac{\tilde{\chi}}{a}=0,\label{constraintPSI}\\
Z -\left(1-\frac{3}{2}\frac{A}{m^2}\right) \fracc{\sigma}{a}=0,\label{constraintZ}\\
\left(\frac{2}{3} - \frac{B}{m^2}\right)\frac{X}{a} -\frac{1}{3}\rho\tilde{\chi}-aI-\frac{B}{a}\psi=0.\label{constraintX}
\end{eqnarray}
From these equations, it is possible to write $\psi$ as a function of $\tilde{\chi}$ and $X$ (for $\gamma_{ef}\neq 0$). Besides, from Eq.\ (\ref{constraintZ}) we see that $Z$ is given only in terms of $\sigma$, which lead us to a closed dynamical system in three variables, from which all others are determined.

The qualitative analysis we are interested in can be done through the characteristic polynomial of the system. For a sake of simplicity, we choose a dust dominate era and write this 3-dimensional closed system in conformal time. Then, it yields

\begin{eqnarray}
p(x)=x^3 + \frac{6(B\eta^4 + 4\eta^2 + 12)}{(B\eta^4 + 12)\eta}\, x^2 &-&\frac{1}{3}\frac{(A - 2C)\eta^6m^2 + 18A\eta^4 - (36C + 432)\eta^2 + 216}{\eta^2(B\eta^4 + 12)}\, x +\nonumber\\&-& \frac{2(A\eta^4m^2 - 12\eta^2m^2 + 18A\eta^2 - 12C\eta^2 + 72)}{(B\eta^4 + 12)\eta}.
\end{eqnarray}

Once the system is non-autonomous we study the time evolution of the eigenvalues, i.e. the roots of $p(x)$, and their sensibility w.r.t. $A$, $B$ and $C$. A simple calculation shows that in the domain of interest of the conformal time all eigenvalues ($x_1$, $x_2$ and $x_3$) are real with $x_1$ always positive and $x_2$ and $x_3$ always negative. The discriminant never changes sign in this region. Only when we approach the singularity $\eta\rightarrow0$, we obtain a degenerate case where $x_1\rightarrow+\infty$, $x_2\rightarrow-\infty$ and $x_3\rightarrow0^-$. Since we cannot determine explicitly $A$, $B$ and $C$, this study was done for a large window in the parameter space (the same we will use afterwards to study the growth of perturbations).

\subsection{Growth of perturbations in the limit of large wavelengths}

In order to describe the structure formation, we can also use the local decomposition in irreducible parts of the projected covariant derivative of $\chi_\mu$ as
\begin{equation}
ah_\mu{}^{\lambda} h_\nu{}^{\epsilon}\chi_{\lambda}{}_{;\epsilon}=\frac{1}{3}h_{\mu\nu}\Delta+\Sigma_{\mu\nu}+W_{\mu\nu},
\end{equation}
where $W_{\mu\nu}$ gives the anti-symmetric part, $\Sigma_{\mu\nu}$ is the symmetric traceless part and, according to \cite{bruni92}, the variable $\Delta$ is the scalar gauge invariant variable that represents the clumping of matter. The equation for $\Delta$ can be derived from Eq.\ (\ref{chi}) and up to first order reads
\begin{equation}
\label{deltadot}
\dot{\Delta}=\frac{a^2}{\rho_t}h^{\alpha\beta}(\pi_{\mu\nu}\sigma^{\mu\nu})_{,\alpha;\beta}-\gamma_{ef}ah^{\alpha\beta}Z_{\alpha;\beta}+a^2\gamma_{ef}\theta h^{\alpha\beta}a_{\alpha;\beta}.
\end{equation}

Once the shear viscosity was widely studied previously, in this paper we focus our analysis only on the role of the anisotropic pressure, reflected by the wave number dependent coefficients $A$, $B$ and $C$. Hence, we set $\xi=0$ from now on.

For large wavelengths, that is, small values of $m$, Eq.\ (\ref{sigmat}) gives us $\dot\sigma=-X$. Deriving this equation once more, changing it to conformal time $d\eta=dt/a(t)$ and using Eq.\ (\ref{xt}) yields
\begin{equation}
\sigma^{\prime\prime}+2\frac{a^\prime}{a}\sigma^\prime+\left(C-\frac{1}{2}\gamma_{ef}a^2\rho_t\right)\sigma=0,
\end{equation}
whose solution for a dust dominated phase ($\gamma_{ef}=1$ and $a\propto \eta^{2}$) is
\begin{equation}
\label{larges}
\sigma(\eta) = \frac{c_1}{\eta^{3/2}} {\rm J}\left(\frac{\sqrt{33}}{2},\sqrt{C}\eta\right) +
\frac{c_2}{\eta^{3/2}} {\rm Y}\left(\frac{\sqrt{33}}{2},\sqrt{C}\eta\right),
\end{equation}
where ${\rm J}$ and ${\rm Y}$ are Bessel functions of first and second kind, respectively\footnote{We assume $C>0$, otherwise we would have modified Bessel functions.}, and $c_1$, $c_2$ and $C$ depend on $m$. For a dust-dominated phase the contribution of the PMF to the energy density is negligible, but the anisotropic pressure component, which goes with $1/a^2$, is still important.

Using Eq.\ (\ref{eq_vinc_quant_cine1}) and the expansion $\delta\Delta=\chi(\eta)Q(x^i)$, we get from Eq.\ (\ref{deltadot})
\begin{equation}
\chi'(\eta) = -\frac{Am^2}{\rho\,a}\,\sigma(\eta) - \frac{3\gamma_{ef}m^2}{2}\,\left( \frac{2}{3}\,a -\frac{B}{a m^2} \right)\sigma(\eta),
\end{equation}
where $A$ and $B$ depend on $m$ and the divergence of the acceleration that is proportional to $m^2$ has been neglected. In the conformal time, we see that $A$, $B$ and $C$ have units of inverse of wavelength squared [$m^2$]. Therefore, using the limit of small values of the argument in (\ref{larges}), $\sqrt{C}\eta\ll1$, we explicitly obtain
\begin{equation}
\label{chisol}
\chi(\eta)=\frac{c_1}{\Gamma\left(\frac{\sqrt{33}}{2}\right)}\left[\frac{3(\sqrt{33}+5)}{8}\,B+\frac{(3-\sqrt{33})m^2\,\eta^4}{12\,\eta_0{}^4} + \frac{(\sqrt{33} - 7)A m^2\,\eta^6}{96\,\eta_0{}^4} \right] \frac{\left(\sqrt{C} \eta \right)^{\frac{\sqrt{33}}{2}}}{(\eta/\eta_0)^{5/2}}
\end{equation}
where we have neglected the decaying mode by setting $c_2=0$.

\subsection{Comparison with the Friedmann model}
\label{comparison}

\begin{figure}[!htb]
\centering
\includegraphics[width=10cm,height=9cm]{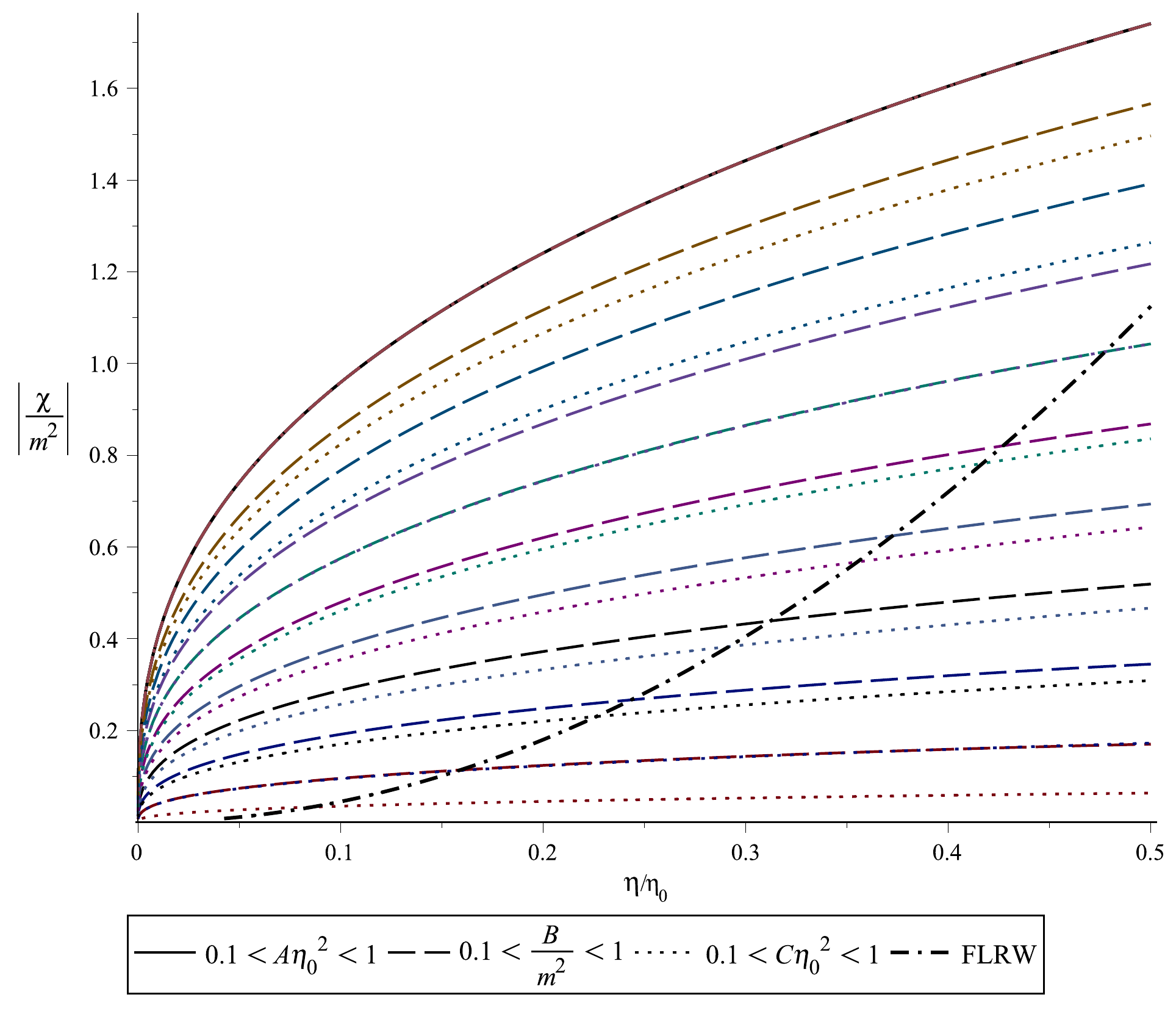}
\caption{(colors online) Growth of perturbations for different values of the quantities $A$, $B$ and $C$ compared to the dust dominated FLRW case. We have set $c_1=1$ in all cases. Note that the curves related to variations in $A\eta_0^2$ (with $B/m^2$ and $C\eta_0^2$ set to unity) remain very close to each other, whereas variations in $B/m^2$ (with $A\eta_0^2$ and $C\eta_0^2$ set to unity) and variations of $C\eta_0^2$ (with $A\eta_0^2$ and $B/m^2$ set to unity) have a larger range of variation such that the bigger the value of these parameters the faster the growth rate.}
\label{fig1}
\end{figure}

The equivalent of Eq. (\ref{chisol}) in the case of a matter dominated FLRW universe is given by \cite{novello}

\begin{equation}
\chi(\eta)=\frac{c_1}{6}\,\left(\frac{\eta}{\eta_0}\right)^2
\end{equation}
We compare the growth in time of the quantity above with the modified version, Eq. (\ref{chisol}), in Fig.\ (\ref{fig1}).
We have plotted different curves $\chi(\eta)$ normalized by $m^2$ for different values of the dimensionless quantities $A\eta_0^2$, $B/m^2$ and $C\eta_0^2$ together with the standard matter-dominated FLRW. Note that even though $A$, $B$ and $C$ are not completely independent and are ultimately related to the constant $k$ present in the metric (\ref{fried_r_pi}), we vary them as if they were to get an intuitive view of how they affect the evolution of $\chi$. We clearly see that variations of one order of magnitude in $A\eta_0^2$, with $B/m^2$ and $C\eta_0^2$ held fixed (set to unity), produce small variations on the solution, while variations of $B/m^2$ and $C\eta_0^2$ are more significant in the sense that they produce larger variations in the comoving fractional energy density gradient when varied in one order of magnitude. 

This analysis shows us that it is possible to obtain a larger growth of the inhomogeneities when compared to a dust-dominated FLRW, in the limit of large wavelengths, for a very large range of the parameters. The full analysis of the dependence of the solution on these parameters and consequently on the wavenumber $m$ of course demands the exact solution of Eq.(\ref{r_eq}) and this requires a much deeper analysis that is not in the scope of this work. In spite of that, we see that the Weyl tensor might be efficient for gravitational collapse by allowing a larger growth rate of the perturbations and could {\it in principle} mimic in part the dark matter component in structure formation in this scenario.

\section{Concluding Remarks}

We have performed a perturbative analysis of a quasi-Friedmann model with a non-null Weyl tensor directly related to an anisotropic pressure component due to a primordial magnetic field. We have adopted the covariant approach to perturbations and suitable gauge-invariant variables directly related to observational quantities were used. We have shown that, depending on the values of the parameters involved, it is possible to have a larger growth of the perturbations when compared to the standard FLRW model, which could in principle play the role of dark matter or at least diminish its contribution in structure formation. Clearly, an analysis of the dependence on the wavenumber of the constants involved is needed to confirm it and in particular to treat the issue of scale invariance of the perturbations and the asymptotic behavior for small scales, which shall be treated in a future work. Anyhow, it is a very interesting fact that taking into account a non-null anisotropic pressure component which is hidden at the non-perturbative level could give us a scenario where perturbations are enhanced and that might not fully rely on an unknown dark component.

\acknowledgments The authors EB and GBS acknowledge the financial support provided by the CAPES-ICRANet program through the grants BEX 13956/13-2 and 13955/13-6.

\appendix

\section{Appendix: The Quasi-Maxwellian equations}\label{appendix}

We know that a Riemannian geometry satisfies the Bianchi identities. In particular, in the case of general relativity, the Bianchi identities together with the Einstein equations yield the Quasi-Maxwellian equations of gravity. These equations are easily obtained if we rewrite the Riemann tensor in terms of its traces and the Weyl tensor, according to the relations defined in section \ref{notation}.

Assuming Einstein constant equal to $1$, the Bianchi identities become
\begin{equation}
\label{bianchi}
W^{\alpha\beta\mu\nu}{}_{;\nu}=-\fracc{1}{2}T^{\mu[\alpha;\beta]}+
\fracc{1}{6}g^{\mu[\alpha}T^{,\beta]},
\end{equation}
where the square brackets mean anti-symmetrization.

The reason of the nomenclature is due to several analogies between the Quasi-Maxwellian and the Maxwell equations. However, this similarity does not go further because the QM equations are in fact highly non-linear and of higher order of differentiability in comparison to Maxwell's theory, leading to situations that never happen in the last case. Indeed, the similitude appears when we make the projection of the QM equations with respect to the vector field $V^{\alpha}$ and its orthogonal hypersurface. At this point, it is very useful to replace the Weyl tensor by its electric $E_{\alpha\beta}$ and magnetic $H_{\alpha\beta}$ parts:

\begin{equation}
\nonumber
\begin{array}{lcl}
E_{\alpha\beta}&\doteq&-W_{\alpha\mu\beta\nu}V^{\mu}V^{\nu},\\[2ex]
H_{\alpha\beta}&\doteq&-^{*}W_{\alpha\mu\beta\nu}V^{\mu}V^{\nu},
\end{array}
\end{equation}
where $^{*}W_{\alpha\mu\beta\nu}$ is the dual of Weyl tensor constructed with the skew-symmetric Levi-Civita tensor. In the same way as the Faraday tensor $F_{\mu\nu}$, the Weyl tensor can be written in terms of its electric and magnetic parts, as follows

\begin{eqnarray}
\nonumber
W_{\alpha\beta}{}^{\mu\nu} = 2\, V_{[\alpha}E_{\beta]}{}^{[\mu}V^{\nu]} + \delta_{[\alpha}^{[\mu}E_{\beta]}^{\nu]} - \eta_{\alpha\beta\lambda\sigma}V^{\lambda}H^{\sigma[\mu}V^{\nu]}+\\
-\eta^{\mu\nu\lambda\sigma}V_{\lambda}H_{\sigma[\alpha}V_{\beta]}.\nonumber
\end{eqnarray}

Considering the kinematical quantities defined in Sec.\ \ref{notation}, the four independent projections of the Bianchi identities

\begin{equation}
\label{proj_div_weyl}
\begin{array}{l}
W^{\alpha\beta\mu\nu}{}_{;\nu}V_{\beta}V_{\mu}h_{\alpha}{}^{\sigma},\\[1ex]
W^{\alpha\beta\mu\nu}{}_{;\nu}\eta^{\sigma\lambda}{}_{\alpha\beta}V_{\mu}V_{\lambda},\\[1ex]
W^{\alpha\beta\mu\nu}{}_{;\nu}h_{\mu}{}^{(\sigma}\eta^{\tau)\lambda}{}_{\alpha\beta}V_{\lambda},\\[1ex]
W^{\alpha\beta\mu\nu}{}_{;\nu}V_{\beta}h_{\mu(\tau}h_{\sigma)\alpha},
\end{array}
\end{equation}
lead to the following independent equations

\begin{equation}
\label{quase_max1}
\begin{array}{l}
h^{\epsilon\alpha}h^{\lambda\gamma}E_{\alpha\lambda;\gamma} + \eta^{\epsilon}{}_{\beta\mu\nu}V^{\beta}H^{\nu\lambda}\sigma^{\mu}{}_{\lambda} + 3H^{\epsilon\nu}\omega_{\nu} = \fracc{1}{3}h^{\epsilon\alpha}\rho_{,\alpha} + \, \fracc{\theta}{3}q^{\epsilon} -\fracc{1}{2}(\sigma^{\epsilon}{}_{\nu} - 3\omega^{\epsilon}{}_{\nu})q^{\nu} + \\[2ex]
\fracc{1}{2}\pi^{\epsilon\mu}a_{\mu} + \fracc{1}{2}h^{\epsilon\alpha}\pi_{\alpha}{}^{\nu}{}_{;\nu};
\end{array}
\end{equation}

\begin{equation}
\label{quase_max2}
\begin{array}{l}
h^{\epsilon\alpha}h^{\lambda\gamma}H_{\alpha\lambda;\gamma} - \eta^{\epsilon}{}_{\beta\mu\nu}V^{\beta}E^{\nu\lambda}\sigma^{\mu}{}_{\lambda} - 3E^{\epsilon\nu}\omega_{\nu} = (\rho+p)\omega^{\epsilon} - \,\fracc{1}{2}\eta^{\epsilon\alpha\beta\lambda}V_{\lambda}q_{\alpha;\beta} + \\[2ex]
\fracc{1}{2}\eta^{\epsilon\alpha\beta\lambda}(\sigma_{\mu\beta} + \omega_{\mu\beta})\pi^{\mu}{}_{\alpha}V_{\lambda};
\end{array}
\end{equation}

\begin{equation}
\label{quase_max3}
\begin{array}{l}
h_{\mu}{}^{\epsilon}h_{\nu}{}^{\lambda}\dot H^{\mu\nu} + \theta H^{\epsilon\lambda} - \fracc{3}{2}H_{\alpha}{}^{(\epsilon}\sigma^{\lambda)\alpha}+H^{\alpha\beta}\sigma_{\alpha\beta}h^{\epsilon\lambda} -a_{\alpha}E_{\beta}{}^{(\lambda}\eta^{\epsilon)\gamma\alpha\beta}V_{\gamma}+ \fracc{1}{2}E_{\beta}{}^{\mu}{}_{;\alpha} h_{\mu}^{(\epsilon}\eta^{\lambda)\gamma\alpha\beta}V_{\gamma} =\\[2ex]
- \fracc{3}{4}q^{(\epsilon}\omega^{\lambda)}
+ \fracc{1}{2}h^{\epsilon\lambda}q^{\mu}\omega_{\mu}
+\fracc{1}{4}\sigma_{\beta}{}^{(\epsilon}\eta^{\lambda)\alpha\beta\mu}V_{\mu}q_{\alpha} +\fracc{1}{4}h^{\nu(\epsilon}\eta^{\lambda)\alpha\beta\mu}V_{\mu}\pi_{\nu\alpha;\beta};
\end{array}
\end{equation}

\begin{equation}
\label{quase_max4}
\begin{array}{l}
h_{\mu}{}^{\epsilon}h_{\nu}{}^{\lambda}\dot E^{\mu\nu} + \theta E^{\epsilon\lambda} - \fracc{3}{2}E_{\alpha}{}^{(\epsilon}\sigma^{\lambda)\alpha}+E^{\alpha\beta}\sigma_{\alpha\beta}h^{\epsilon\lambda} +a_{\alpha}H_{\beta}{}^{(\lambda}\eta^{\epsilon)\gamma\alpha\beta}V_{\gamma} - \fracc{1}{2}H_{\beta}{}^{\mu}{}_{;\alpha} h_{\mu}^{(\epsilon}\eta^{\lambda)\gamma\alpha\beta} V_{\gamma} = \\[2ex] -\fracc{1}{2}(\rho+p)\sigma^{\epsilon\lambda} +\fracc{1}{6}h^{\epsilon\lambda}(q^{\mu}{}_{;\mu} -q^{\mu}a_{\mu} -\pi^{\mu\nu}\sigma_{\mu\nu})
+\fracc{1}{2}q^{(\epsilon}a^{\lambda)} -\fracc{1}{4}h^{\mu(\epsilon}h^{\lambda)\alpha}q_{\mu;\alpha} + \fracc{1}{2}h_{\alpha}{}^{\epsilon}h_{\mu}{}^{\lambda}\dot\pi^{\alpha\mu} + \\[2ex] \fracc{1}{4}\pi_{\beta}{}^{(\epsilon}\sigma^{\lambda)\beta}- \fracc{1}{4}\pi_{\beta}{}^{(\epsilon}\omega^{\lambda)\beta} + \fracc{1}{6}\theta\pi^{\epsilon\lambda}.
\end{array}
\end{equation}
These are the Quasi-Maxwellian equations and it is clear the similitude to the Maxwell equations: the first two correspond to $\nabla\cdot \vec E$ and $\nabla\cdot\vec H$, while the the last pair gives the time evolution of $\vec H$ and $\vec E$, respectively. To obtain a self-consistent system of equations we need to add the energy-momentum tensor conservation law $T^{\mu\nu}{}_{;\nu}=0$, which gives
\begin{equation}
\label{expl_proj_conserv_mom_eneg1}
\dot\rho + (\rho+p)\theta + \dot q^{\mu}V_{\mu} + q^{\alpha}{}_{;\alpha} - \pi^{\mu\nu}\sigma_{\mu\nu} = 0,
\end{equation}
and
\begin{equation}
\label{expl_proj_conserv_mom_eneg2}
(\rho+p)a_{\alpha} - p_{\mu}h^{\mu}{}_{\alpha} + \dot q_{\mu}h^{\mu}{}_{\alpha} + \theta q_{\alpha} + q^{\nu}\theta_{\alpha\nu} + q^{\nu}\omega_{\alpha\nu} + h_{\alpha}{}^{\beta}\pi_{\beta}{}^{\nu}{}_{;\nu} = 0.
\end{equation}
The integrability condition

$$V^{\alpha}{}_{;\mu;\nu}-V^{\alpha}{}_{;\nu;\mu}=R^{\alpha}{}_{\beta\mu\nu}V^{\beta}$$
applied to the observer field chosen, can be translated into evolution equations plus constraints for the kinematical quantities. Thereby, the evolution equations are
\begin{equation}
\label{evol_quant_cine1}
\dot\theta + \fracc{\theta^2}{3} + 2(\sigma^2 + \omega^2) - a^{\alpha}{}_{;\alpha}= -\fracc{1}{2}(\rho+3p),
\end{equation}

\begin{equation}
\label{evol_quant_cine2}
\begin{array}{l}
h_{\alpha}{}^{\mu}h_{\beta}{}^{\nu}\dot\sigma_{\mu\nu} + \fracc{1}{3}h_{\alpha\beta}(a^{\lambda}{}_{;\lambda}-2\sigma^2-2\omega^2) + a_{\alpha}a_{\beta}-\fracc{1}{2}h_{\alpha}{}^{\mu}h_{\beta}{}^{\nu}a_{(\mu;\nu)} + \fracc{2}{3}\theta\sigma_{\alpha\beta} + \sigma_{\alpha\mu}\sigma^{\mu}{}_{\beta}+\\[2ex]
\omega_{\alpha\mu}\omega^{\mu}{}_{\beta} =  -E_{\alpha\beta} - \fracc{1}{2}\pi_{\alpha\beta},
\end{array}
\end{equation}

\begin{equation}
\label{evol_quant_cine3}
h_{\alpha}{}^{\mu}h_{\beta}{}^{\nu}\dot\omega_{\mu\nu} - \fracc{1}{2}h_{\alpha}{}^{\mu}h_{\beta}{}^{\nu}a_{[\mu;\nu]} +\fracc{2}{3}\theta\omega_{\alpha\beta} -\sigma_{\beta\mu}\omega^{\mu}{}_{\alpha}+\sigma_{\alpha\mu}\omega^{\mu}{}_{\beta}=0,
\end{equation}
and the constraint equations are

\begin{equation}
\label{eq_vinc_quant_cine1}
\fracc{2}{3}\theta_{,\mu}h^{\mu}{}_{\lambda} - (\sigma^{\alpha}{}_{\gamma} + \omega^{\alpha}{}_{\gamma})_{;\alpha}h^{\gamma}{}_{\lambda} - a^{\nu}(\sigma_{\lambda\nu} + \omega_{\lambda\nu}) =-q_{\lambda},
\end{equation}

\begin{equation}
\label{eq_vinc_quant_cine2}
\omega^{\alpha}{}_{;\alpha}+2\omega^{\alpha}a_{\alpha}=0,
\end{equation}

\begin{equation}
\label{eq_vinc_quant_cine3}
H_{\tau\lambda} = -\fracc{1}{2} h_{(\tau}{}^{\epsilon} h_{\lambda)}{}^{\alpha} \eta_{\epsilon}{}^{\beta\gamma\nu} V_{\nu}(\sigma_{\alpha\beta} + \omega_{\alpha\beta})_{;\gamma} + a_{(\tau}\omega_{\lambda)}.
\end{equation}

According to Lichnerowicz's theorem \cite{lich60}, the complete set of equations presented above propagates the solutions of the Einstein equations defined only on a given Cauchy surface to the whole space-time.

We remark that under our assumptions--the background metric being written in a synchronous coordinate system, the cosmological fluid being represented by $\rho$, $p$ and $\pi_{\mu\nu}$ and its comoving observer being geodesic, shear-free and irrotational--it is easy to see that Eqs.\ (\ref{quase_max2}, \ref{evol_quant_cine3}-\ref{eq_vinc_quant_cine3}) are identically zero. Moreover, the spherical symmetry of the model suggests that the anisotropic pressure tensor can be put in the form $[\pi^{\mu}{}_{\nu}]=f(t)g(r)\,\mbox{diag}(-1,1/2,1/2),$ where $f(t)$ and $g(r)$ are arbitrary functions to be determined (see Sec. \ref{background}). In this case, the non-trivial equations are exactly Eqs.\ (\ref{qm_gfried}) and Eq.\ (\ref{quase_max3}) is then identically verified.

\end{document}